# Efficient sampling for Gaussian linear regression with arbitrary priors

P. Richard Hahn *

Arizona State University

and

Jingyu He

Booth School of Business, University of Chicago

and

Hedibert F. Lopes

INSPER Institute of Education and Research

May 24, 2018

## Abstract

This paper develops a slice sampler for Bayesian linear regression models with arbitrary priors. The new sampler has two advantages over current approaches. One, it is faster than many custom implementations that rely on auxiliary latent variables, if the number of regressors is large. Two, it can be used with any prior with a density function that can be evaluated up to a normalizing constant, making it ideal for investigating the properties of new shrinkage priors without having to develop custom sampling algorithms. The new sampler takes advantage of the special structure of the linear regression likelihood, allowing it to produce better effective sample size per second than common alternative approaches.

---
*The authors gratefully acknowledge the Booth School of Business for support. The third author acknowledges the support of research fellowships from the CNPq and FAFESP, Brazil.

# 1   Introduction

This paper develops a computationally efficient posterior sampling algorithm for Bayesian linear regression models with Gaussian errors. Our new approach is motivated by the fact that existing software implementations for Bayesian linear regression do not readily handle problems with large number of observations (hundreds of thousands) and predictors (thousands). Moreover, existing sampling algorithms for popular shrinkage priors are bespoke Gibbs samplers based on case-specific latent variable representations. By contrast, the new algorithm does not rely on case-specific auxiliary variable representations, which allows for rapid prototyping of novel shrinkage priors outside the conditionally Gaussian framework.

Specifically, we propose a slice-within-Gibbs sampler based on the elliptical slice sampler of Murray et al. [2010]. These authors focus on sampling from posteriors that are proportional to a product of a multivariate Gaussian prior and an arbitrary likelihood. Intuitively, the elliptical slice sampler operates by drawing samples from the Gaussian factor of the posterior and then accepting or rejecting these samples by evaluating the non-Gaussian factor. The starting point of this paper is the observation that the elliptical slice sampler is also suited to the Bayesian Gaussian linear regression case, which has a multivariate Gaussian likelihood and an arbitrary prior (that is, the roles of the likelihood and prior are reversed). In fact, under independent priors over the regression coefficients, the Gaussian likelihood term contains all of the co-dependence information of the posterior, allowing us to pre-compute many key quantities, leading to a remarkably efficient, generic algorithm.

After explaining the new sampler in detail in the following section, extensive computational demonstrations are provided in Section 3. The new sampling approach is demonstrated on the horseshoe prior [Carvalho et al., 2010], the Laplace prior [Park and Casella, 2008, Hans, 2009] and the independent Gaussian or "ridge" prior. These three priors boast widely-available, user-friendly implementations. Although other shrinkage priors have been proposed and studied, many have not been implemented in the regression setting and hence are not widely used outside of the normal-means context; consequently, we restrict our comparison to three popular regression priors. Recently developed priors that are not considered here include the Dirichlet-Laplace prior [Bhattacharya et al., 2015], the normal-gamma prior [Caron and Doucet, 2008, Griffin et al., 2010, Griffin and Brown, 2011], the Bayesian bridge [Polson et al., 2014] and many others [Armagan, 2009, Armagan et al., 2011, 2013, Neville et al., 2014, Polson and Scott, 2010].

We further demonstrate the flexibility of our approach by using it to perform posterior inference under two non-standard, "exotic", priors — an asymmetric Cauchy prior and a "non-local" two-component mixture prior — for which there exist no standard samplers. Our approach is also suitable for the case where there are more predictors than observations; in this regime, we compare to Johndrow and Orenstein [2017] whose new Markov chain Monte Carlo algorithm works for tens of thousands of variables. Our method is implemented in `C++` as the package `bayeslm` [Hahn et al., 2018] in `R` [R Core Team, 2017].

## 2   Elliptical slice sampling for shrinkage regression with arbitrary priors

### 2.1   Review of elliptical slice sampling for Gaussian priors

To begin, we review the elliptical slice sampler of Murray et al. [2010]. In the following subsections we adapt the sampler specifically for use with Gaussian linear regression models. Unless otherwise noted, random variables (possibly vector-valued) are denoted by capital Roman letters, matrices are in bold, vectors are in Roman font, and scalars are italic. All vectors are column vectors.

The elliptical slice sampler considers cases where the goal is to sample from a distribution $p(\Delta) \propto N(\Delta; 0, \mathbf{V})L(\Delta)$. The key idea is to take advantage of the elementary fact that the sum of two Gaussian random variables is a Gaussian random variable. Accordingly, for two independent (vector) random variables $v_0 \sim N(0, \mathbf{V})$ and $v_1 \sim N(0, \mathbf{V})$ and for any $\theta \in [0, 2\pi]$, $\Delta = v_0 \sin \theta + v_1 \cos \theta$ is also distributed according to $N(0, \mathbf{V})$, since $\text{cov}(\Delta) = \mathbf{V} \sin^2 \theta + \mathbf{V} \cos^2 \theta = \mathbf{V}$. Because this holds for each $\theta$, the marginal distribution of $\Delta$ is $N(0, \mathbf{V})$ for any distribution over $\theta$.

Therefore, Murray et al. [2010] note that if one can sample from the parameter-expanded model $p(v_0, v_1, \theta) \propto \pi(\theta)N(v_0; 0, \mathbf{V})N(v_1; 0, \mathbf{V})L(v_0 \sin \theta + v_1 \cos \theta)$, then samples from $p(\Delta)$ can be obtained as samples of the transformation $v_0 \sin \theta + v_1 \cos \theta$. Sampling from this model is easiest to explain in terms of a singular Gaussian prior distribution over $(v_0^t, v_1^t, \Delta^t)^t$ with covariance

$$\boldsymbol{\Sigma}_\theta = \begin{pmatrix} \mathbf{V} & 0 & \mathbf{V}\sin\theta \\ 0 & \mathbf{V} & \mathbf{V}\cos\theta \\ \mathbf{V}\sin\theta & \mathbf{V}\cos\theta & \mathbf{V} \end{pmatrix}$$

and joint density $p(v_0, v_1, \Delta, \theta) \propto N(0, \boldsymbol{\Sigma}_\theta) L(v_0 \sin\theta + v_1 \cos\theta)$. Using this model, we sample the parameters $(v_0, v_1, \Delta, \theta)$ via a two-block Gibbs sampler:

1. Sample from $p(v_0, v_1 \mid \Delta, \theta)$, which can be achieved by sampling $v \sim N(0, \mathbf{V})$ and setting $v_0 = \Delta \sin\theta + v \cos\theta$ and $v_1 = \Delta \cos\theta - v \sin\theta$.

2. Sample from $p(\Delta, \theta \mid v_0, v_1) \propto N(0, \boldsymbol{\Sigma}_\theta) L(v_0 \sin\theta + v_1 \cos\theta)$ compositionally in two steps:

    (a) First draw from $p(\theta \mid v_0, v_1)$ by marginalizing over $\Delta$, yielding $p(\theta \mid v_0, v_1) \propto L(v_0 \sin\theta + v_1 \cos\theta)$. We draw from this distribution via a traditional one-dimensional slice sampler [Neal, 2003]. Initialize $a = 0$ and $b = 2\pi$.

    i. Draw $\ell$ uniformly on $[0, L(v_0 \sin\theta + v_1 \cos\theta)]$.

    ii. Sample $\theta'$ uniformly on $\theta \in [a, b]$.

    iii. If $L(v_0 \sin\theta' + v_1 \cos\theta') > \ell$, set $\theta \leftarrow \theta'$. Otherwise, shrink the support of $\theta'$ (if $\theta' < \theta$, set $a \leftarrow \theta'$; if $\theta' > \theta$, set $b \leftarrow \theta'$), and go to step ii.

    (b) Then we draw from $p(\Delta \mid \theta, v_0, v_1)$, which is degenerate at $\Delta = v_0 \sin\theta + v_1 \cos\theta$.

Note that this version of the elliptical slice sampler is somewhat different than the versions presented in Murray et al. [2010], but as it reduces to a Gibbs sampler, its validity is more transparent and practically the algorithms are nearly equivalent.

## 2.2 Elliptical slice sampling for Gaussian linear regression

In this section we adapt the sampler described above to permit efficient sampling from Bayesian linear regression models. Specifically, we consider the standard Bayesian linear regression model:

$$Y = \mathbf{X}\beta + \epsilon, \tag{1}$$

where $Y$ is an $n$-by-1 vector of responses, $\mathbf{X}$ is an $n$-by-$p$ matrix of regressors, $\beta$ is a $p$-by-1 vector of coefficients and $\epsilon \sim \mathrm{N}(0, \sigma^2 \mathbf{I})$ is an $n$-by-1 vector of error terms. Denote the prior of $\beta$ as $\pi(\beta)$. The objective is to sample from a posterior expressible as

$$\pi(\beta \mid \mathrm{y}, \mathbf{X}, \sigma^2) \propto f(\mathrm{y} \mid \mathbf{X}, \beta, \sigma^2 \mathbf{I}) \pi(\beta) \tag{2}$$

where $f(\mathrm{y} \mid \mathbf{X}, \beta)$ denotes a multivariate Gaussian density with mean vector $\mathbf{X}\beta$ and diagonal covariance $\sigma^2 \mathbf{I}$. Our approach is driven by the fact that, up to proportionality, this $n$-dimensional Gaussian can be regarded as the posterior of $\beta$ under a flat prior (indicated by the 0 subscript):

$$\begin{aligned}\pi_0(\beta \mid \sigma^2, \mathrm{y}, \mathbf{X}) \propto {}& \frac{1}{\sqrt{2\pi\sigma^2}} \exp\left[-\frac{1}{2\sigma^2}(\mathrm{y} - \mathbf{X}\widehat{\beta})^T(\mathrm{y} - \mathbf{X}\widehat{\beta})\right] \times \\ & \frac{1}{\sqrt{2\pi\sigma^2}} \exp\left[-\frac{1}{2\sigma^2}(\beta - \widehat{\beta})^T \mathbf{X}^T \mathbf{X}(\beta - \widehat{\beta})\right]\end{aligned} \tag{3}$$

which is the density of a $p$-dimensional Gaussian with mean $\hat{\beta} = (\mathbf{X}^T \mathbf{X})^{-1} \mathbf{X}^T \mathrm{y}$ (the ordinary least squares estimate) and covariance $\sigma^2 (\mathbf{X}^T \mathbf{X})^{-1}$. Therefore, the slice sampler of Murray et al. [2010] can be applied directly, using $\pi_0(\beta \mid \sigma^2, \mathrm{y}, \mathbf{X})$ as the Gaussian "prior" and $\pi(\beta)$ as the "likelihood". One minor modification is that, because $\pi_0(\beta \mid \sigma^2, \mathrm{y}, \mathbf{X})$ is centered at OLS estimator $\widehat{\beta}$, as opposed to 0, we sample the offset of $\beta$ from $\widehat{\beta}$, which we denote $\Delta = \beta - \hat{\beta}$.

This sampler is flexible because the only requirement is that the prior function $\pi(\beta)$ can be evaluated up to a normalizing constant. The sampler is efficient, per iteration, because in each Monte Carlo iteration, the sampler draws a single multivariate Gaussian random variable, and then draws from a univariate uniform distribution within the while loop. The size of the sampling region for $\theta$ shrinks rapidly with each rejected value and is guaranteed to eventually accept. Sampling of $\sigma^2$ can be done after sampling $\beta$ in each iteration.

Despite being quite fast per iteration, for larger regression problems, with $p$ having more than a few dozen elements, the autocorrelation from this joint sampler can be prohibitively high, yielding very low effective sample sizes. Intuitively, this occurs because for any given auxiliary variables $(v_0, v_1)$, the slice step over $\theta$ frequently has only a very narrow acceptable region, entailing that subsequent samples of $\theta$ (and hence $\beta$) will be very close to one another. Fortunately, the basic strategy of the elliptical slice sampler can be applied to smaller blocks, an approach we describe in the following section.

**Algorithm 1** : *Elliptical slice sampler for linear regression*

For initial value $\beta$, with $\Delta = \beta - \hat{\beta}$, and $\sigma^2$ fixed:

1. Draw $v \sim \mathrm{N}(0, \sigma^2(\mathbf{X}^T\mathbf{X})^{-1})$. Set $v_0 = \Delta \sin\theta + v\cos\theta$ and $v_1 = \Delta\cos\theta - v\sin\theta$.

2. Draw $\ell$ uniformly on $[0, \pi(\hat{\beta} + v_0\sin\theta + v_1\cos\theta)]$. Initialize $a = 0$ and $b = 2\pi$.

    (a) Sample $\theta'$ uniformly on $[a, b]$.

    (b) If $\pi(\hat{\beta} + v_0\sin\theta' + v_1\cos\theta') > \ell$, set $\theta \leftarrow \theta'$ and go to step 3. Otherwise, shrink the support of $\theta'$ (if $\theta' < \theta$, set $a \leftarrow \theta'$; if $\theta' > \theta$, set $b \leftarrow \theta'$), and go to step 2(a).

3. Return $\Delta = v_0\sin\theta + v_1\cos\theta$ and $\beta = \hat{\beta} + \Delta$.

Figure 1: The elliptical slice sampler for linear regression (with an arbitrary prior) samples all $p$ elements of the regression coefficients simultaneously.

## 2.3 Elliptical slice-within-Gibbs for linear regression

As mentioned above, if the number of regression coefficients $p$ is large, the slice which contains acceptable proposals is likely to be minuscule. Due to the shrinking bracket mechanism of the slice sampler, it rejects many proposals and shrinks the bracket strongly towards the initial value, thereby inducing high autocorrelation in the obtained samples. Here, we propose a slice-within-Gibbs sampler (Figure 2) to mitigate this problem.

Because $\beta$ has a jointly Gaussian likelihood and independent priors, it is natural to implement a Gibbs sampler, updating a subset, which we denote $\beta^k$, given all other coefficients, which we denote $\beta^{-k}$ (and other parameters), in each MCMC iteration. This is possible because the conditional distribution for the Gaussian portion of the distribution, which accounts for all the dependence, is well-known and easy to sample from. The basic idea is simply to apply the sampler from Figure 1 using the conditional distribution $\beta^k \mid \beta^{-k}$ as the "likelihood" instead of the full likelihood $L(\beta)$.

From equation 3, the joint likelihood of $\beta$ is $N(\widehat{\beta}, \sigma^2(\mathbf{X}^T\mathbf{X})^{-1})$. Therefore, we group elements of $\beta$ into several blocks $\beta = (\beta_1, \cdots, \beta_p) = \{\beta^1, \beta^2, \cdots, \beta^K\}$, constructing a Gibbs sampling scheme for all $K$ blocks, using the elliptical slice sampler to update each block. We can rearrange terms of the joint distribution as

$$\begin{bmatrix} \beta^k \\ \beta^{-k} \end{bmatrix} \sim N\left( \begin{bmatrix} \widehat{\beta}^k \\ \widehat{\beta}^{-k} \end{bmatrix}, \sigma^2 \begin{bmatrix} \mathbf{\Sigma}_{k,k} & \mathbf{\Sigma}_{k,-k} \\ \mathbf{\Sigma}_{-k,k} & \mathbf{\Sigma}_{-k,-k} \end{bmatrix} \right) \tag{4}$$

where $\begin{bmatrix} \widehat{\beta}^k \\ \widehat{\beta}^{-k} \end{bmatrix} = \widehat{\beta}$, the OLS estimator and $\begin{bmatrix} \mathbf{\Sigma}_{k,k} & \mathbf{\Sigma}_{k,-k} \\ \mathbf{\Sigma}_{-k,k} & \mathbf{\Sigma}_{-k,-k} \end{bmatrix} = (\mathbf{X}^T\mathbf{X})^{-1}$.

The corresponding conditional distribution of $\beta^k$ given $\beta^{-k}$ is $N(\tilde{\beta}^k, \tilde{\mathbf{\Sigma}}^k)$ where

$$\tilde{\beta}^k = \widehat{\beta}^k + \mathbf{\Sigma}_{k,-k}\mathbf{\Sigma}_{-k,-k}^{-1}(\beta^{-k} - \widehat{\beta}^{-k}) \tag{5}$$

$$\tilde{\mathbf{\Sigma}}^k = \sigma^2 \left( \mathbf{\Sigma}_{k,k} - \mathbf{\Sigma}_{k,-k}\mathbf{\Sigma}_{-k,-k}^{-1}\mathbf{\Sigma}_{-k,k} \right). \tag{6}$$

Note that the grouping of coefficients is arbitrary; if all coefficients are grouped in a single block, we recover the original sampler from Figure 1. Empirically, we find that putting each coefficient in a different block so that $K = p$, and updating coefficients one by one, gives excellent performance. Our complete algorithm is given in Figure 2.

**Algorithm 2** : *Slice-within-Gibbs sampler for linear regression*

- For each $k$ from 1 to $K$

  Update $\beta^k \mid \beta^{\{-k\}}, \sigma^2, \lambda$ according to Algorithm 1.

  1. Construct the conditional mean $\tilde{\beta}^k$ and conditional covariance matrix $\tilde{\Sigma}^k$ as in expressions (5) and (6). Set $\Delta^k = \beta^k - \tilde{\beta}^k$. Draw $v \sim N(0, \tilde{\Sigma}^k)$. Set $v_0 = \Delta^k \sin\theta^k + v\cos\theta^k$ and $v_1 = \Delta^k \cos\theta^k - v\sin\theta^k$.
  2. Draw $\ell$ uniformly on $[0, \pi(\Delta^k + \tilde{\beta}^k)]$. Initialize $a = 0$ and $b = 2\pi$.
     (a) Sample $\theta'$ uniformly on $[a, b]$.
     (b) If $\pi(\tilde{\beta}^k + v_0 \sin\theta' + v_1 \cos\theta') > \ell$, set $\theta^k \leftarrow \theta'$. Otherwise, shrink the support of $\theta'$ (if $\theta' < \theta^k$, set $a \leftarrow \theta'$; if $\theta' > \theta^k$, set $b \leftarrow \theta'$), and go to step (a).
  3. Return $\Delta^k = v_0 \sin\theta^k + v_1 \cos\theta^k$ and $\beta^k = \tilde{\beta}^k + \Delta^k$.

- Update $\sigma^2 \mid \boldsymbol{\beta}, \lambda$: let $s = (\mathrm{y} - \mathbf{X}\beta)^T(\mathrm{y} - \mathbf{X}\beta)$. Draw $\sigma^2 \sim \mathrm{IG}((n+\alpha)/2, (s+\gamma)/2)$, where IG denotes the inverse-gamma distribution and $(\alpha, \gamma)$ are the associated prior parameters.

- Update $\lambda \mid \boldsymbol{\beta}, \sigma^2$ via a random walk Metropolis-Hastings step on the log scale with a diffuse Gaussian prior:

  1. Draw $r \sim \mathrm{N}(0, 0.2^2)$, let $\lambda_{\text{proposal}} = \exp(\log(\lambda) + r)$.
  2. Compute the Metropolis-Hastings ratio

  $$\eta = \exp(\log\pi(\beta, \lambda_{\text{proposal}}) - \log\pi(\beta, \lambda) + \log(\lambda_{\text{proposal}}) - \log(\lambda)) \qquad (7)$$

  3. Draw $u \sim \mathrm{Unif}(0,1)$, if $u < \eta$, accept $\lambda_{\text{proposal}}$; otherwise keep the current $\lambda$.

Figure 2: The full slice-within-Gibbs sampler, including update steps for $\sigma^2$ and $\lambda$.

## 2.4 Computational considerations

Although the slice-within-Gibbs sampler updates the coefficients iteratively, it can be more efficient than Gibbs samplers based on conditionally Gaussian representations because the structure of the slice sampler allows the necessary matrix factorizations and inversions to be pre-computed outside the main Gibbs loop. Specifically, to efficiently compute the $K$ conditional mean vectors and covariance matrices given in expressions (5) and (6) we can precompute $\mathbf{\Sigma}_{k,-k}\mathbf{\Sigma}_{-k,-k}^{-1}$, $\mathbf{\Sigma}_{k,k} - \mathbf{\Sigma}_{k,-k}\mathbf{\Sigma}_{-k,-k}^{-1}\mathbf{\Sigma}_{-k,k}$, and Cholesky factors $\mathbf{L}_k$, with $\mathbf{L}_k\mathbf{L}_k^T = \mathbf{\Sigma}_{k,k} - \mathbf{\Sigma}_{k,-k}\mathbf{\Sigma}_{-k,-k}^{-1}\mathbf{\Sigma}_{-k,k}$, for each $k = 1, \ldots, K$. By contrast, Gibbs samplers based on conditionally Gaussian representations (e.g., Armagan et al. [2011]) have full conditional updates of the form

$$(\beta \mid \sigma^2, \mathbf{D}) \sim \mathrm{N}((\mathbf{X}^T\mathbf{X} + \mathbf{D})^{-1}\mathbf{X}^T\mathrm{y}, \sigma^2(\mathbf{X}^T\mathbf{X} + \mathbf{D})^{-1}), \qquad (8)$$

which require costly Cholesky or eigenvalue decompositions of the matrix $(\mathbf{X}^T\mathbf{X} + \mathbf{D})^{-1}$ at each iteration as $\mathbf{D}$ is updated — eliminating this step at each iteration is the primary savings of the new algorithm. From this basic observation two notable facts follow:

1. Because the efficiency of our sampler relies on precomputing these statistics, it is not immediately applicable to regression models that require data augmentation at the observation level, such as models allowing non-Gaussian errors via mixtures of normals (e.g. $t$-distributed errors as scale-mixture of normals or location mixtures for multi-modal error distributions). For such expanded models, the analogous expressions involve latent parameters that vary across sampling iterations. It is possible that our slice sampler could be modified for non-Gaussian errors, but the extension would not be straightforward and we do not consider it further here.

2. Parallelization improves the traditional Gibbs sampler more than it improves the ellipitical slice sampler. Note that computation of (8) can be improved with parallelization of linear algebra routines, meaning that this improvement is realized in each iteration. By contrast, the slice sampler can parallelize the pre-computation of the conditional distributions, but this is a one-time upfront benefit. Accordingly, comparisons between the new slice sampling approach and the traditional samplers will depend on whether or not parallelization is utilized, such as via the efficient Math Kernel Library (MKL) linear algebra library. Some of our simulations use MKL and some do not, but we clearly indicate in each caption whether parallelization has been employed. Figure 3 shows the impact of additional processors on the performance comparison using effective sample size per second (as defined in Section 3.3).

Finally, although the slice approach may incur additional computational cost if many proposals are rejected in each iteration prior to acceptance, we find that not to be the case. Figure 4 plots the average number of rejections before one accepted draw against average running time under different signal-to-noise ratios. When signal-to-noise ratio is high, the likelihood is strong, thus the elliptical slice sampler rejects fewer proposals and is accordingly faster. For signal-to-noise ratios between 1 and 4, the expected number of rejections is approximately constant.

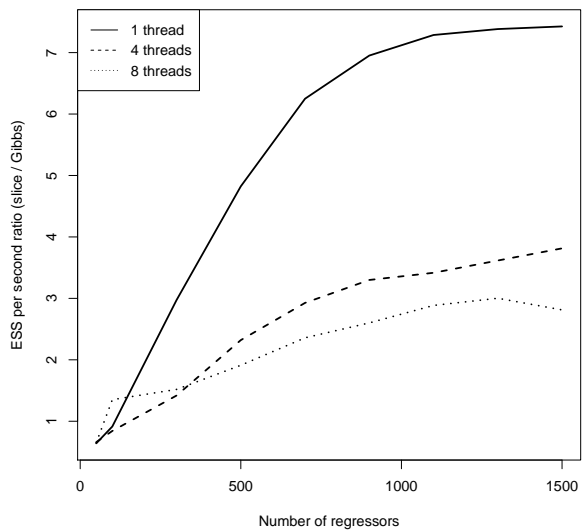

Figure 3: Ratio of effective sample size per second of the elliptical slice sampler and `monomvn` Gibbs sampler with different numbers of regressors and numbers of threads used in parallel. The ratio increases, indicating that the elliptical slice sampler is faster, with the number of regressors, but decreases as the number of threads increases because the MKL library improves the performance of the `monomvn` package. However, the elliptical slice sampler is still much faster than Gibbs sampler.

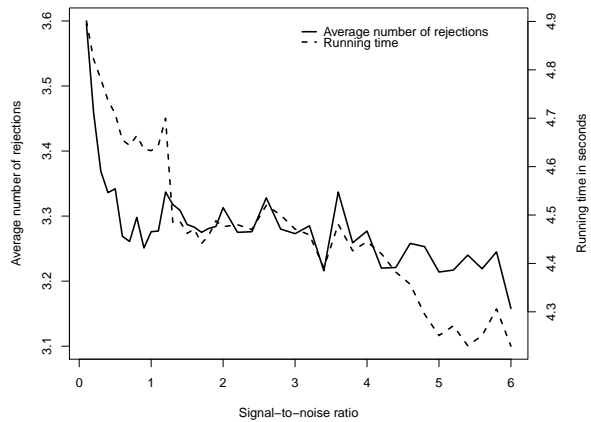

Figure 4: Average number of rejections in each iteration prior to acceptance and raw computing time over a range of signal-to-noise ratios. Here we show the result of horseshoe regression with $p = 100$ and $n = 1000$, drawing $12,000$ posterior samples.

## 2.5 The rank-deficient case

It is increasingly common to want to analyze regression models with more predictors than observations: $p > n$. Similarly, it is sometimes the case that $\mathbf{X}^T\mathbf{X}$ can be rank-deficient due to perfect collinearity in the predictor matrix $\mathbf{X}$. Such cases would seem to pose a problem for our method for the following reason. Recall that the slice sampler draws from a target distribution of the form

$$\begin{aligned} p(\beta \mid \mathrm{y}, \mathbf{X}, \sigma) &\propto \mathrm{N}_Y(\mathbf{X}\beta, \sigma^2)\pi(\beta) \\ &\propto \mathrm{N}_\beta(\hat{\beta}, \sigma^2(\mathbf{X}^T\mathbf{X})^{-1})\pi(\beta), \end{aligned} \qquad (9)$$

where we abuse notation somewhat and use $\mathrm{N}(\cdot, \cdot)$ to denote the Gaussian distribution function. If $\mathrm{rank}(\mathbf{X}^T\mathbf{X}) = r < p$, then the first term on the right-hand side is not absolutely continuous with respect to the second term, and the sampler will not function properly. Intuitively, the proposal distribution is supported on an $r < p$ dimensional hyperplane. The slice sampler will never propose values off of this subspace; hence it cannot have the correct target distribution. Operationally, $\hat{\beta}$ is not even unique. Fortunately, the algorithm can be salvaged with a very minor modification inspired by ridge regression analysis. We rewrite the above expression as

$$
\begin{aligned}
p(\beta \mid y, \mathbf{X}, \sigma) &\propto \mathrm{N}_Y(\mathbf{X}\beta, \sigma^2)\mathrm{N}_\beta(0, c\sigma^2\mathbf{I})\frac{\pi(\beta)}{\mathrm{N}_\beta(0, c\sigma^2\mathbf{I})}, \\
&\propto \mathrm{N}_\beta(\bar{\beta}, \sigma^2(\mathbf{X}^T\mathbf{X} + c^{-1}\mathbf{I})^{-1})\frac{\pi(\beta)}{\mathrm{N}_\beta(0, c\sigma^2\mathbf{I})},
\end{aligned}
\tag{10}
$$

for $c > 0$, where $\bar{\beta} = (\mathbf{X}^T\mathbf{X} + c^{-1}\mathbf{I})^{-1}\mathbf{X}^T y$. In the first line we merely "multiplied by 1"; subsequent lines reorganize the distributions in the familiar form required by the slice sampler. This reformulation solves the problem of absolute continuity; $\bar{\beta}$ is now well-defined and $(\mathbf{X}^T\mathbf{X} + c^{-1}\mathbf{I})$ is full rank. Otherwise, the algorithm operates exactly as before, only feeding in $\bar{\beta}$ rather than $\hat{\beta}$ and $(\mathbf{X}^T\mathbf{X} + c^{-1}\mathbf{I})$ rather than $\mathbf{X}^T\mathbf{X}$ and evaluating the ratio $\pi(\beta)/\mathrm{N}(0, c\sigma^2\mathbf{I})$ where one would have otherwise evaluated the prior $\pi(\beta)$. The optimal value of $c$ will differ depending on the data and the prior being used; in practice small values near one seem to work fine, but tuning based on pilot runs could be performed if desired.

## 3 Simulation studies

In this section we compare the performance of our new algorithm against several well-known alternatives. Specifically, we apply our approach to the horseshoe prior [Carvalho et al., 2010], the Laplace prior [Park and Casella, 2008, Hans, 2009] and the independent Gaussian or "ridge" prior. These three priors are frequently used in empirical studies in part because they have readily available implementations.

The goal here is merely to demonstrate the efficacy of our computational approach, not to advocate for any particular prior choice. Indeed, our hope is that having a generic sampling scheme for any prior will make computational considerations secondary when choosing one's prior. Ideally, one would not select a prior merely on the grounds that it admits an efficient sampling algorithm. In other words, the selling point of the present approach is not that it is strictly better than the existing samplers for these models (it is not necessarily), rather it is that we are using the *same* underlying algorithm for all three of them, with no custom modifications, and are still achieving competitive (or superior) computational performance. In the $p > n$ regime, we also compare to the recent algorithm of Johndrow and Orenstein [2017]; default Gibbs samplers were too costly time-wise to conduct simulations for $p > 500$. In the following subsections, we detail the priors considered as well as our data-generating process.

## 3.1 Priors

We investigate three standard priors: the horseshoe prior [Carvalho et al., 2010], the Laplace prior [Park and Casella, 2008, Hans, 2009], and the conjugate Gaussian prior (ridge regression). Details of these priors are provided here for reference.

*Horseshoe prior.* The horseshoe prior can be expressed as a local scale-mixture of Gaussians

$$\beta \sim N(0, \lambda^2 \Lambda^2), \quad \lambda \sim C^+(0,1), \quad \lambda_1, ..., \lambda_p \stackrel{\text{iid}}{\sim} C^+(0,1), \tag{11}$$

where $C^+(0,1)$ is a half standard Cauchy distribution, $\Lambda = \text{diag}(\lambda_1, ...\lambda_p)$ represents the local shrinkage parameters and $\lambda$ is the global shrinkage parameter. The standard approach to sampling from the posterior of regressions under horseshoe priors is a Gibbs sampler which samples $(\lambda_1, \ldots, \lambda_p)$ from their full conditionals.

The horseshoe density, integrating over the local scale factors $\lambda_j$, can be computed using special functions. However, the following bounds [Carvalho et al., 2010] provide an excellent approximation which is more straightforward to evaluate:

$$\frac{1}{2\sqrt{2\pi^3}} \log\left(1 + \frac{4}{(\beta_j/\lambda)^2}\right) < \pi(\beta_j/\lambda) < \frac{1}{\sqrt{2\pi^3}} \log\left(1 + \frac{2}{(\beta_j/\lambda)^2}\right). \tag{12}$$

In our implementation we use the lower bound as our prior density function.

*Laplace prior.* The Laplace (double-exponential) prior is given by

$$\pi(\beta_j \mid \lambda) = \frac{1}{2} \lambda^{-1} \exp(-|\beta_j|/\lambda). \tag{13}$$

Park and Casella [2008] gives the first treatment of Bayesian lasso regression and Hans [2009] proposes alternative Gibbs samplers.

*Ridge prior.* The ridge prior is given by

$$\beta \mid \lambda \sim N(0, \lambda^2 \mathbf{I}_p). \tag{14}$$

Section 2.3.2 of Gamerman and Lopes [2006] provides a nice exposition of the general Bayesian linear and Gaussian regression model.

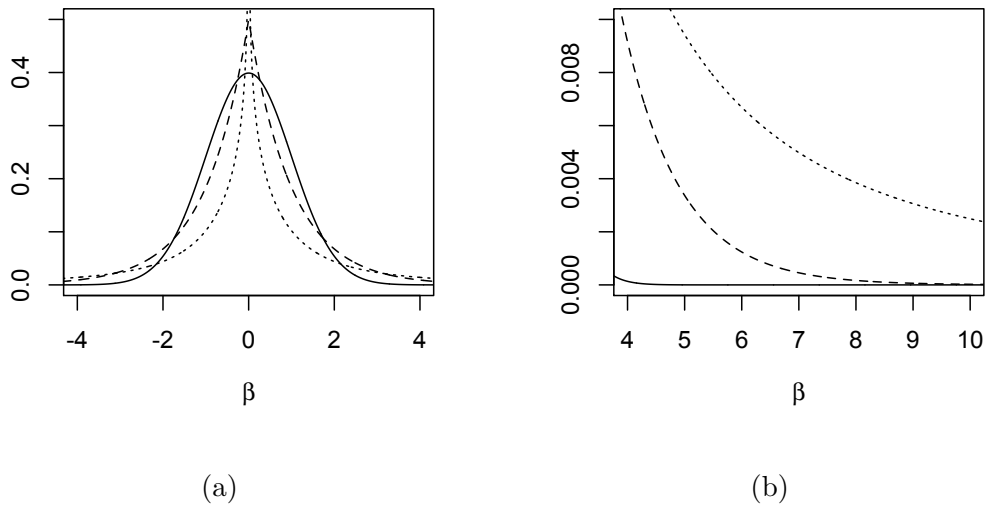

Figure 5: Three priors for regression coefficients: Gaussian (solid), Laplace (dashed), horseshoe (dotted) when $\lambda = 1$. Panel (a) shows the body of the distributions and panel (b) shows a zoomed-in look at the tails.

Note that all three priors have a "global" shrinkage parameter $\lambda$, which is given a hyperprior. A key feature of Bayesian shrinkage regression is the inference of this parameter; as opposed to setting it at a fixed value or selecting it by cross-validation, point estimates of $\beta$ are obtained as the marginal posterior mean, integrating over (the posterior of) $\lambda$. Figure 5 plots these three densities for comparison.

## 3.2 Data generating process

The predictor matrix $\mathbf{X}$ is generated in two different ways: independently from a standard Gaussian distribution, or according to a Gaussian factor model so that variables have strong linear codependencies. Details of the data-generating process are shown below.

1. Draw elements of $\beta$ from a "sparse Gaussian" where $\lceil p \rceil$ entries of $\beta$ are non-zero,

drawn from a standard Gaussian distribution, and all other entries are zero.

2. Generate the regressors matrix $\mathbf{X}$ in one of two ways.

    - In the independent regressor case, every element of the regressor matrix $\mathbf{X}_{n \times p}$ is drawn independently from a standard Gaussian distribution.

    - In the factor structure case, we draw each row of $\mathbf{X}$ from a Gaussian factor model with $k = p/5$ factors. Latent factor scores are drawn according to $\mathbf{F}_{k \times n} \sim$ N(0, 1). The factor loading matrix, $\mathbf{B}_{p \times k}$, has entries that are either zero or one, with exactly five ones in each column and a single 1 in each row, so that $\mathbf{BB}^t$ is block diagonal, with blocks of all ones and all other elements being zero. The regressors are then set as $\mathbf{X} = (\mathbf{BF})^T + \boldsymbol{\varepsilon}$ where $\boldsymbol{\varepsilon}$ is a $n \times p$ matrix of errors with independent N(0, 0.01) entries.

3. Set $\sigma = \kappa \sqrt{\sum_{j=1}^{p} \beta_j^2}$, where $\kappa$ controls noise level.

4. Draw $y_i = \mathbf{x}_i' \beta + \epsilon_i, \epsilon_i \sim \mathrm{N}(0, \sigma^2)$ for $i = 1, \ldots, n$.

Additionally, we vary the noise level, letting $\kappa = 1$ or $\kappa = 2$, corresponding to signal-to-noise ratios of 1 and 1/2, respectively.

## 3.3 Comparison metrics

To gauge the performance of our new algorithm, we must judge not only the speed, but also the quality of the posterior samples. To address this concern, we compare our approach with alternative samplers using effective sample size per second (see e.g. Gamerman and Lopes [2006] pages 126 - 127). Letting $N$ denote the Monte Carlo sample size, the effective sample size $N_{\text{eff}}(\beta_j)$ is

$$N_{\text{eff}}(\beta_j) = \frac{N}{1 + 2 \sum_{k=1}^{\infty} \rho_k}, \tag{15}$$

where $\rho_k = \mathrm{corr}\left(\beta_j^{(0)}, \beta_j^{(k)}\right)$ is the autocovariance of lag $k$. To verify that the samplers are giving comparable results (they ought to be fitting the same model) we also report the estimation error of the associated posterior point estimates. Suppose $\{\bar{\beta}_j\}$ are posterior means of each variable and $\{\beta_j\}$ are true values. The estimation error is measured by

$$\text{error} = \sqrt{\frac{\sum_{j=1}^{p}(\bar{\beta}_j - \beta_j)^2}{\sum_{j=1}^{p} \beta_j^2}}. \tag{16}$$

Although we do not report it here, we also examined posterior standard deviations and found all algorithms to be comparable up to Monte Carlo error. For each simulation, 50,000 posterior samples are drawn, 20,000 of which are burn-in samples (with no thinning). We divide $N_{\text{eff}}$ by running time in seconds to compute ESS per second as a measure of efficiency of each sampler.

## 3.4 Simulation study results

The `R` package `monomvn` [Gramacy, 2017] implements the standard Gibbs samplers for the horseshoe prior (function `bhs`), the Laplace prior (function `blasso`), and Gaussian prior (function `bridge`) prior. For the Laplace prior, we additionally compare with the Gibbs sampler from Hans [2009]. All of the samplers are implemented in `C++`. Tables 1 and 2 report a representative subset of our simulation results; comprehensive tables can be found in the Appendix. Here we summarize the broad trends that emerge.

First, the slice sampler enjoys a substantial advantage in terms of effective sample size (ESS) per second compared to the standard samplers. For example, in the independent regressor case (Table 1), when $p = 1,000$ and $n = 10 \times p$, our approach is about 18 times faster than the `monomvn` Gibbs sampler.

When there exist strong colinearities in the regressor matrix (Table 2) the new approach, which samples one coefficient at a time, loses some of its efficiency compared to the standard algorithms, which draw the regression coefficients jointly. However, the new approach is still superior when $p > 1000$.

In addition to effective sample size per second, we also consider raw computing time. The code is tested on a machine with an Intel i7-6920HQ CPU and 16GB RAM. For the horseshoe regression with independent regressors, with $p = 500$ and $n = 5,000$, the slice-within-Gibbs sampler takes 101 seconds running time to draw 50,000 posterior samples, of which 19 seconds are fixed computing time and 82 seconds are spent within the loop. By comparison, the standard Gibbs sampler takes 1 second of fixed computing time and 5,310

|  |  |  | Error |  |  |  | ESS per second |  |  |
|---|---|---|---|---|---|---|---|---|---|
| Prior | $p$ | $n$ | OLS | slice | monomvn | Gibbs | slice | monomvn | Gibbs |
| Horseshoe | 100 | 10p | 3.38% | 1.52% | 1.51% | 1.51% | 1399 | 613 | 567 |
|  | 1000 | 10p | 1.05% | 0.27% | 0.27% | 0.27% | 91 | 5 | 5 |
| Laplace | 100 | 10p | 3.38% | 2.39% | 2.38% | – | 2362 | 809 | – |
|  | 1000 | 10p | 1.04% | 0.63% | 0.63% | – | 168 | 8 | – |
| Ridge | 100 | 10p | 3.38% | 3.20% | 3.20% | – | 3350 | 959 | – |
|  | 1000 | 10p | 1.06% | 0.99% | 0.99% | – | 178 | 5 | – |

Table 1: Simulation results of all three priors, where $\kappa = 1$ and all regressors are independent. The table demonstrates error and effective sample size per second of the elliptical slice sampler, Gibbs sampler in R package monomvn (column monomvn ) and our own implementation of Gibbs sampler for horseshoe regression (column Gibbs). The elliptical slice sampler has similar error to the Gibbs sampler, but much higher effective sample size per second.

seconds within the loop to draw the same number of posterior samples.

### 3.4.1 The $p > n$ setting

In the $p > n$ regime we compare our algorithm to the sampler presented in Johndrow and Orenstein [2017][1]. This choice was based on the fact that the standard Gibbs samplers are prohibitively slow for $p = 1000$, with a runtime of 1450 seconds versus 92 seconds when $n = 900$. Indeed, the infeasibility of existing methods for an applied data set with $p = 1500$ and $n = 300,000$ was our initial motivation for developing the approach described in this paper. The Johndrow and Orenstein [2017] approach was developed contemporaneously to our method, but focuses on the case where $p \gg n$ (such as genome-wide association studies).

First, we note that both methods give similar root mean squared error (RMSE), suggesting that the posteriors being sampled from are comparable. Second, the elliptical slice sampler scales well with $n$ given a fixed $p$. For example, the ESS per second of our approach is around 40 when $p = 1000$ and $n$ ranges from 300 to 900. By contrast, Johndrow and Orenstein [2017] does well when $p \gg n$ such as when $p = 3000$ and $n = 100$; in that case its ESS per second is as high as 69. However, the ESS per second drops significantly when $n$ becomes larger for their approach. A full comparison is displayed in Table 3.

---
[1] We are grateful to the authors for making their Matlab code available.

| Prior | $p$ | $n$ | Error | | | | ESS per second | | |
|---|---|---|---|---|---|---|---|---|---|
| | | | OLS | 1-block | monomvn | Gibbs | 1-block | monomvn | Gibbs |
| Horseshoe | 100 | 10p | 16.47% | 6.06% | 6.04% | 6.03% | 387 | 747 | 792 |
| | 1000 | 10p | 6.85% | 1.64% | 1.64% | 1.64% | 36 | 4 | 4 |
| Laplace | 100 | 10p | 17.06% | 7.21% | 7.15% | – | 573 | 1257 | – |
| | 1000 | 10p | 6.77% | 1.95% | 1.94% | – | 38 | 5 | – |
| Ridge | 100 | 10p | 16.90% | 8.50% | 8.75% | – | 669 | 1668 | – |
| | 1000 | 10p | 6.85% | 2.93% | 3.09% | – | 38 | 6 | – |

Table 2: Simulation results of all three priors, $\kappa = 1$. The regressors are not independent but have underlying factor structure, in that every five regressors are highly correlated with one another. The elliptical slice sampler has similar error to the Gibbs sampler, but much higher effective sample size per second when $p = 1,000$.

# 4 Empirical illustration: beauty and course evaluations

In this section, we consider an interesting data set first presented in Hamermesh and Parker [2005]. The data are course evaluations from the University of Texas at Austin between 2000 and 2002. The data are on a 1 to 5 scale, with larger numbers being better. In addition to the course evaluations, information concerning the class and the instructor were collected. To quote Hamermesh and Parker [2005]:

> We chose professors at all levels of the academic hierarchy, obtaining professorial staffs from a number of departments that had posted all faculty members' pictures on their departmental websites. An additional ten faculty members' pictures were obtained from miscellaneous departments around the University. The average evaluation score for each undergraduate course that the faculty member taught during the academic years 2000-2002 is included. This sample selection criterion resulted in 463 courses, with the number of courses taught by the sample members ranging from 1 to 13. The classes ranged in size from 8 to 581 students, while the number of students completing the instructional ratings ranged from 5 to 380. Underlying the 463 sample observations are 16,957 completed evaluations from 25,547 registered students.

|  |  |  | Running Time | | RMSE | | ESS per second | |
| --- | --- | --- | --- | --- | --- | --- | --- | --- |
| $p$ | $n$ | $\kappa$ | J&O | Slice | J&O | Slice | J&O | Slice |
| 1000 | 300 | 0.25 | 119.11 | 91.50 | 0.0041 | 0.0038 | 46.71 | 43.19 |
| 1000 | 600 | 0.25 | 394.02 | 88.61 | 0.0028 | 0.0026 | 14.68 | 47.26 |
| 1000 | 900 | 0.25 | 905.36 | 88.91 | 0.0021 | 0.0020 | 6.60 | 48.85 |
| 1000 | 300 | 1 | 127.33 | 90.19 | 0.0189 | 0.0189 | 43.92 | 39.25 |
| 1000 | 600 | 1 | 399.50 | 91.17 | 0.0129 | 0.0129 | 14.39 | 44.12 |
| 1000 | 900 | 1 | 927.96 | 91.58 | 0.0098 | 0.0099 | 6.35 | 46.09 |
| 1500 | 450 | 0.25 | 346.37 | 187.91 | 0.0029 | 0.0027 | 16.37 | 21.26 |
| 1500 | 900 | 0.25 | 1073.28 | 185.57 | 0.0022 | 0.0021 | 5.50 | 23.08 |
| 1500 | 1350 | 0.25 | 2629.52 | 183.68 | 0.0018 | 0.0017 | 2.27 | 24.04 |
| 1500 | 450 | 1 | 326.63 | 183.66 | 0.0164 | 0.0164 | 17.39 | 20.28 |
| 1500 | 900 | 1 | 1021.47 | 174.52 | 0.0100 | 0.0101 | 5.73 | 23.72 |
| 1500 | 1350 | 1 | 2515.37 | 176.51 | 0.0071 | 0.0071 | 2.36 | 24.78 |
| 3000 | 100 | 0.25 | 85.95 | 985.68 | 0.0067 | 0.0075 | 69.72 | 3.89 |
| 3000 | 500 | 0.25 | 575.92 | 983.64 | 0.0024 | 0.0022 | 9.85 | 4.17 |

Table 3: Comparison of effective sample size (ESS) per second and root mean squared error (RMSE) with Johndrow and Orenstein [2017] (denoted J&O) for the $p > n$ case. Both samplers take $12,000$ posterior draws with the first $2,000$ as burn-in. Both samplers give similar RMSE. The elliptical slice sampler has higher ESS per second in most cases considered here, especially when $p \approx n$. The Johndrow et al. sampler is much more efficient only when $p \gg n$, such as $p = 3000$ and $n = 100$. Results were tabulated *without* the MKL linear algebra library.

For additional details on how the beauty scores were constructed and on how to interpret the regression results, see Hamermesh and Parker [2005]. Here we do not aim to provide any sort of definitive reanalysis of the results in Hamermesh and Parker [2005]. Instead, our goal is to fit a plausible, but over-parametrized, model to their data and to employ a variety of priors, including some non-standard priors in addition to the usual shrinkage priors (horseshoe, Laplace and ridge). We are interested in finding out whether conclusions change substantively under "exotic" priors that are not likely to be used by the typical social scientist.

The model we fit allows for fixed effects for each of 95 instructors[2]. We include additive effects for the following factors: class size (number of students), language in which the professor earned his or her degree, whether or not the instructor was a minority, gender, beauty rating, and age. Each of these variables was included in the model via dummy variables according to the following breakdown:

- class size: below 31, 31 to 60, 61 to 150, or 151 to 600 (four levels by quartile),
- language: English or non-English (two levels),
- minority: ethnic minority or non-minority (two levels),
- gender: male or female (two levels),
- beauty: four levels by quartile, and

---

[2] We recover the instructors by matching on teacher characteristics, including a variable denoting if the professor's photo is in black and white or in color. Using this method we find 95 uniques, although the original paper says there are 94 instructors. The data we used can be found at http://faculty.chicagobooth.edu/richard.hahn/teaching/hamermesh.txt

- age: below 43, 43 to 47, 48 to 56 and 57 to 73 (four levels by quartile).

Finally, we include up to three-way interactions between age, beauty, and gender. We include an intercept in our model so that individual effects can be interpreted as deviation from the average. Three predictors are then dropped because they are collinear with the intercept (and each other); there were no highly beautiful males between 42 and 46, and no moderately beautiful instructors of either gender in that same age group. Even after dropping these columns, the model is numerically singular, with only 98 out of 130 singular values greater than 1e-14.

In addition to the horseshoe, Laplace and ridge regression priors, we also analyze these data using two exotic priors: an asymmetric Cauchy prior and a "non-local" two-component mixture of Cauchys.

The asymmetric Cauchy prior is

$$\pi(\beta) = \begin{cases} 2qf(\beta) & \beta \leq 0 \\ 2f(\beta/s)(1-q)/s & \beta > 0 \end{cases}, \tag{17}$$

where $f(x) = \frac{1}{\pi(1+x^2)}$ is the density of the standard Cauchy distribution and $s = (1-q)/q$. Here, $q$ is the prior probability that the coefficient is negative. We refer to this prior as having a shark fin density, as suggested by the shape shown in Figure 6. The bivariate mixture of Cauchys prior is

$$\pi(\beta) = 0.5t(\beta; -1.5, 1) + 0.5t(\beta; 1.5, 1) \tag{18}$$

where $t(x; m, v)$ is the density of the Student-$t$ distribution with location $m$ and degrees of freedom $v$. The non-local mixture of Cauchys is a sort of "anti-sparsity" prior: it asserts that the coefficients are all likely to be similar in magnitude and non-zero, but with unknown sign. A global scale parameter can be accommodated within the above forms by using density $\pi(\beta/\lambda)/\lambda$. Figure 6 displays the density functions of these two priors.

When applying the exotic priors to the course evaluations model, we differentiated between regressors in terms of hyperparameter selection. Specifically, in the asymmetric Cauchy model we defaulted to $q = 0.5$, except for the following: for the largest class size, we set $q = 0.75$, for tenure track status we set $q = 0.25$, for non-English we set $q = 0.75$,

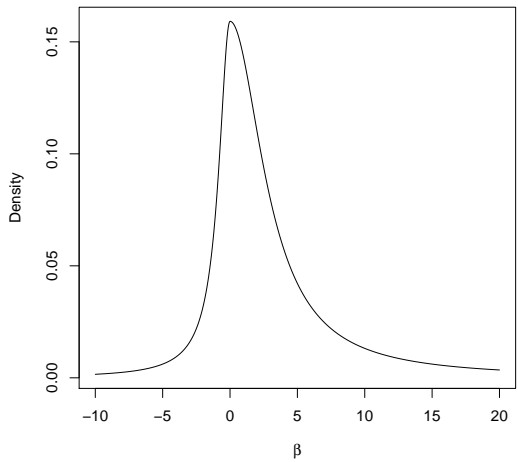 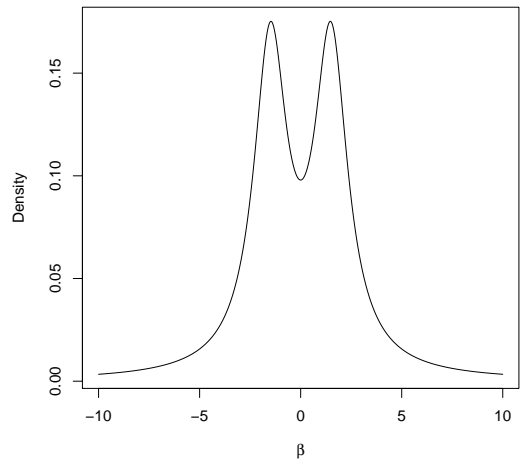

(a) Density of sharkfin density.

(b) Density of mixture of Cauchy, location parameter $-1.5$ and $1.5$.

Figure 6: Panel (a) depicts the density of the "shark fin" prior with $q = 0.25$. Panel (b) depicts the density of a two-component Cauchy mixture distribution where the weight for each component is $1/2$, the scale parameter is 1, and the location parameters are $-1.5$ and $1.5$, respectively.

and for each of the three higher beauty levels we set $q = 0.25$. Similarly, for the non-local Cauchy mixture, we defaulted to a standard Cauchy, using the non-local prior only for the class size, tenure track, language, and minority variables.

Our results are summarized in Table 4, which reports all variables whose posterior 95% quantile-based symmetric credible interval excluded zero for at least one of the five priors. Figure 7 shows kernel density plots of posterior samples of these coefficients. All posterior inferences were based on effective sample sizes of approximately 20,000. A number of features stand out. First, there is a relatively small set of factors that are isolated between the various models as statistically significant (in the Bayesian sense described above); this suggests that the data is meaningfully overwhelming the contributions of the priors. Likewise, we note that the signs on the point estimates concur across all five priors. Second, we note that different priors do make a difference, both in terms of which variables among this set are designated significant and also in terms of the magnitude of the point estimates obtained. Third, one specific difference in the results that is noteworthy is that the horseshoe prior does not flag beauty as significant, while the ridge prior gives a much larger estimate. In Figure 7, the horseshoe posterior (black line) appears notably different than the other priors in panel (e), exhibiting bimodality, where one mode is at 0 and another mode is located away from zero in the direction of the maximum likelihood estimate; this distinctive shape is consistent with the horseshoe prior's aggressive shrinkage profile at the origin.

Table 4: Posterior points estimates of regression coefficients under each prior; those whose posterior 95% credible intervals exclude zero are shown in bold.

| variable name | horseshoe | lasso | ridge | sharkfin | non-local |
|---|---|---|---|---|---|
| class size 61 to 150 | $-0.13$ | $\mathbf{-0.19}$ | $\mathbf{-0.20}$ | $\mathbf{-0.14}$ | $\mathbf{-0.22}$ |
| class size 151 to 600 | $\mathbf{-0.36}$ | $\mathbf{-0.41}$ | $\mathbf{-0.43}$ | $\mathbf{-0.36}$ | $\mathbf{-0.46}$ |
| tenure track | $0.22$ | $\mathbf{0.29}$ | $0.31$ | $\mathbf{0.27}$ | $\mathbf{0.40}$ |
| non-minority | $\mathbf{0.65}$ | $\mathbf{0.65}$ | $\mathbf{0.53}$ | $\mathbf{0.63}$ | $\mathbf{0.71}$ |
| highly beautiful | $0.14$ | $\mathbf{0.36}$ | $\mathbf{0.54}$ | $0.25$ | $\mathbf{0.38}$ |

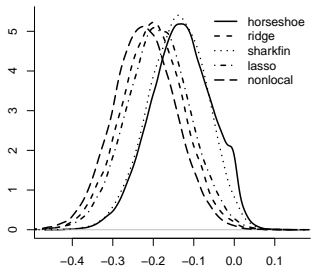
(a) class size 61 to 150

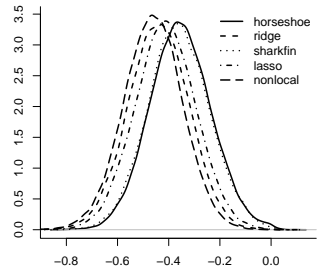
(b) class size 151 to 600

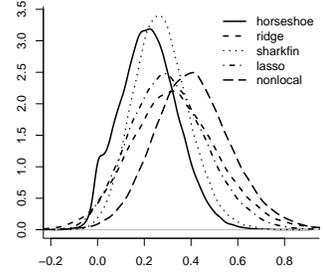
(c) tenure track

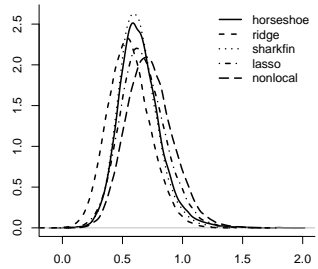
(d) non-minority

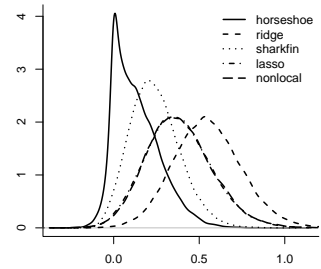
(e) highly beautiful

Figure 7: Kernel density plots of posterior samples of regression coefficients.

# 5 Discussion

This paper presents a new and efficient sampler for the purpose of Bayesian linear regression with arbitrary priors. The new method is seen to be competitive with, or better than, the usual Gibbs samplers that are routinely used to fit such models using popular shrinkage priors. The new approach is flexible enough to handle any class of priors admitting density evaluations. We hope that our new sampling approach will foster research into interesting classes of priors that do not have obvious latent variable representations and to encourage empirical researchers to conduct more bespoke sensitivity analysis.

# A  Complete simulation results

In the appendix we show complete simulation results. The data-generating process is presented in section 3.2. We implement the elliptical slice sampler for horseshoe, Laplace and ridge priors, and compare with the `monomvn` package for all three priors, Hans [2009] for the Laplace prior, and our own implementation of the Gibbs sampler for the horseshoe prior.

## A.1  Independent regressors

In this section, all elements of the regressor matrix $\mathbf{X}_{n\times p}$ are drawn independently from a standard Gaussian distribution. The regression coefficients $\beta$ are drawn from a sparse Gaussian data-generating process as shown in section 3.2. $\kappa$ controls the noise level. All simulations for independent regressors are on machines without the Math Kernel Library (MKL).

### A.1.1 Horseshoe prior

| $p$ | $n$ | Error | | | | ESS per second | | |
|---|---|---|---|---|---|---|---|---|
| | | OLS | slice | monomvn | Gibbs | slice | monomvn | Gibbs |
| 100 | $10p$ | 3.379% | 1.519% | 1.514% | 1.514% | 1399 | 613 | 567 |
| 100 | $50p$ | 1.414% | 0.655% | 0.655% | 0.654% | 1871 | 517 | 670 |
| 100 | $100p$ | 1.007% | 0.464% | 0.464% | 0.463% | 1500 | 301 | 590 |
| 500 | $10p$ | 1.493% | 0.439% | 0.441% | 0.439% | 229 | 29 | 31 |
| 500 | $50p$ | 0.637% | 0.196% | 0.196% | 0.196% | 228 | 20 | 30 |
| 500 | $100p$ | 0.451% | 0.143% | 0.143% | 0.143% | 215 | 10 | 17 |
| 1000 | $10p$ | 1.050% | 0.269% | 0.269% | 0.270% | 91 | 5 | 5 |
| 1000 | $50p$ | 0.448% | 0.113% | 0.113% | 0.113% | 76 | 3 | 3 |
| 1000 | $100p$ | 0.317% | 0.081% | 0.081% | 0.081% | 74 | 2 | 3 |

Table 5: This table compares the error and effective sample size (ESS) per second of various sampling algorithms under the horseshoe prior. The signal-to-noise ratio is $\kappa = 1$, and the response variable is drawn according to the sparse Gaussian model described in the main text. All regressors are mutually independent. Observe that the monomvn package is notably less efficient than our implementation of the Gibbs sampler. We believe this is because monomvn allows $t$-distributed errors, which demands recomputing the sufficient statistics at each iteration, leading it to scale poorly in $n$. This table was generated on a machine *not* running the MKL linear algebra library.

| $p$ | $n$ | Error | | | | ESS per second | | |
|---|---|---|---|---|---|---|---|---|
| | | OLS | slice | monomvn | Gibbs | slice | monomvn | Gibbs |
| 100 | $10p$ | 6.846% | 2.990% | 2.989% | 2.981% | 1248 | 438 | 525 |
| 100 | $50p$ | 2.784% | 1.348% | 1.340% | 1.339% | 1499 | 348 | 588 |
| 100 | $100p$ | 1.989% | 0.884% | 0.883% | 0.883% | 1616 | 331 | 619 |
| 500 | $10p$ | 2.999% | 0.945% | 0.945% | 0.946% | 236 | 21 | 20 |
| 500 | $50p$ | 1.286% | 0.412% | 0.413% | 0.412% | 249 | 15 | 20 |
| 500 | $100p$ | 0.899% | 0.275% | 0.275% | 0.275% | 260 | 12 | 20 |
| 1000 | $10p$ | 2.121% | 0.550% | 0.550% | 0.551% | 79 | 4 | 4 |
| 1000 | $50p$ | 0.912% | 0.245% | 0.245% | 0.245% | 92 | 4 | 6 |
| 1000 | $100p$ | 0.639% | 0.172% | 0.172% | 0.172% | 89 | 3 | 6 |

Table 6: This table compares the error and effective sample size (ESS) per second of various sampling algorithms under the horseshoe prior. The signal-to-noise ratio is $\kappa = 2$, and the response variable is drawn according to the sparse Gaussian model described in the main text. All regressors are mutually independent. Observe that the monomvn package is notably less efficient than our implementation of the Gibbs sampler. We believe this is because monomvn allows $t$-distributed errors, which demands recomputing the sufficient statistics at each iteration, leading it to scale poorly in $n$. This table was generated on a machine *not* running the MKL linear algebra library.

## A.1.2 Laplace prior

|  $p$  |  $n$   | Error  |        |         |        | ESS per second |         |      |
|-------|--------|--------|--------|---------|--------|----------------|---------|------|
|       |        | OLS    | slice  | monomvn | Hans   | slice          | monomvn | Hans |
| 100   | $10p$  | 3.378% | 2.385% | 2.383%  | 2.419% | 2362           | 809     | 1160 |
| 100   | $50p$  | 1.436% | 1.183% | 1.182%  | 1.193% | 3657           | 696     | 1202 |
| 100   | $100p$ | 1.017% | 0.869% | 0.869%  | 0.875% | 3870           | 646     | 1257 |
| 500   | $10p$  | 1.493% | 0.955% | 0.956%  | 0.958% | 472            | 48      | 131  |
| 500   | $50p$  | 0.639% | 0.491% | 0.490%  | 0.495% | 482            | 19      | 149  |
| 500   | $100p$ | 0.447% | 0.364% | 0.364%  | 0.370% | 508            | 16      | 173  |
| 1000  | $10p$  | 1.041% | 0.629% | 0.627%  | 0.631% | 168            | 8       | 78   |
| 1000  | $50p$  | 0.451% | 0.333% | 0.333%  | 0.333% | 156            | 4       | 85   |
| 1000  | $100p$ | 0.317% | 0.249% | 0.249%  | 0.249% | 144            | 3       | 82   |

Table 7: This table compares the error and effective sample size (ESS) per second of various sampling algorithms under a Laplace prior. The signal-to-noise ratio is $\kappa = 1$, and the response variable is drawn according to the sparse Gaussian model described in the main text. All regressors are mutually independent. The ratio column reports the ratio of ESS per second for elliptical slice sampler and that of monomvn package. This table was generated on a machine *not* running the MKL linear algebra library.

| $p$ | $n$ | Error | | | | ESS per second | | |
|---|---|---|---|---|---|---|---|---|
| | | OLS | slice | monomvn | Hans | slice | monomvn | Hans |
| 100 | $10p$ | 6.538% | 3.990% | 3.985% | 3.991% | 2203 | 682 | 1531 |
| 100 | $50p$ | 2.839% | 2.100% | 2.096% | 2.103% | 3034 | 662 | 1487 |
| 100 | $100p$ | 1.966% | 1.539% | 1.537% | 1.537% | 3209 | 370 | 1559 |
| 500 | $10p$ | 2.972% | 1.586% | 1.586% | 1.590% | 421 | 47 | 185 |
| 500 | $50p$ | 1.279% | 0.858% | 0.858% | 0.857% | 398 | 21 | 191 |
| 500 | $100p$ | 0.908% | 0.654% | 0.654% | 0.654% | 524 | 17 | 187 |
| 1000 | $10p$ | 2.105% | 1.057% | 1.054% | 1.056% | 156 | 5 | 63 |
| 1000 | $50p$ | 0.909% | 0.585% | 0.585% | 0.586% | 176 | 4 | 66 |
| 1000 | $100p$ | 0.641% | 0.446% | 0.446% | 0.446% | 171 | 4 | 61 |

Table 8: This table compares the error and effective sample size (ESS) per second of various sampling algorithms under a Laplace prior. The signal-to-noise ratio is $\kappa = 2$, and the response variable is drawn according to the sparse Gaussian model described in the main text. All regressors are mutually independent. The ratio column reports the ratio of ESS per second for elliptical slice sampler and that of monomvn package. This table was generated on a machine *not* running the MKL linear algebra library.

### A.1.3 Ridge regression

| $p$ | $n$ | Error | | | ESS per second | | |
|---|---|---|---|---|---|---|---|
| | | OLS | slice | monomvn | slice | monomvn | ratio |
| 100 | $10p$ | 3.375% | 3.197% | 3.198% | 3350 | 959 | 3.49 |
| 100 | $50p$ | 1.415% | 1.399% | 1.398% | 4714 | 741 | 6.36 |
| 100 | $100p$ | 1.019% | 1.012% | 1.015% | 4720 | 558 | 8.46 |
| 500 | $10p$ | 1.496% | 1.418% | 1.418% | 534 | 28 | 19.07 |
| 500 | $50p$ | 0.642% | 0.636% | 0.636% | 665 | 22 | 30.22 |
| 500 | $100p$ | 0.451% | 0.447% | 0.447% | 645 | 17 | 37.94 |
| 1000 | $10p$ | 1.056% | 0.992% | 0.993% | 178 | 5 | 35.60 |
| 1000 | $50p$ | 0.451% | 0.446% | 0.446% | 213 | 4 | 53.25 |
| 1000 | $100p$ | 0.317% | 0.315% | 0.315% | 228 | 4 | 57 |

Table 9: This table compares the error and effective sample size (ESS) per second of various sampling algorithms under a ridge prior. The signal-to-noise ratio is $\kappa = 1$, and the response variable is drawn according to the sparse Gaussian model described in the main text. All regressors are mutually independent. The ratio column reports the ratio of ESS per second for elliptical slice sampler and that of `monomvn` package. This table was generated on a machine *not* running the MKL linear algebra library.

|   |   | Error | | | ESS per second | | |
| --- | --- | --- | --- | --- | --- | --- | --- |
| $p$ | $n$ | OLS | slice | monomvn | slice | monomvn | ratio |
| 100 | $10p$ | 6.851% | 5.735% | 5.735% | 2810 | 1145 | 2.45 |
| 100 | $50p$ | 2.853% | 2.756% | 2.757% | 4240 | 776 | 5.46 |
| 100 | $100p$ | 2.078% | 2.043% | 2.043% | 4000 | 636 | 6.29 |
| 500 | $10p$ | 2.984% | 2.453% | 2.459% | 516 | 33 | 15.64 |
| 500 | $50p$ | 1.290% | 1.236% | 1.236% | 730 | 26 | 28.08 |
| 500 | $100p$ | 0.892% | 0.875% | 0.875% | 693 | 24 | 28.88 |
| 1000 | $10p$ | 2.098% | 1.722% | 1.726% | 182 | 6 | 30.33 |
| 1000 | $50p$ | 0.904% | 0.867% | 0.868% | 237 | 7 | 35.86 |
| 1000 | $100p$ | 0.642% | 0.629% | 0.629% | 244 | 5 | 48.80 |

Table 10: This table compares the error and effective sample size (ESS) per second of various sampling algorithms. The signal-to-noise ratio is $\kappa = 2$, and the response variable is drawn according to the sparse Gaussian model described in the main text. All regressors are mutually independent. The ratio column reports the ratio of ESS per second for elliptical slice sampler and that of `monomvn` package. This table was generated on a machine *not* running the MKL linear algebra library.

## A.2 Factor structure

In this simulation study, the regressor matrix $\mathbf{X}_{n \times p}$ is drawn with the factor structure shown in section 3.2. All results are generated on machines running the MKL linear algebra library.

| Prior | $p$ | $n$ | Error | | | | ESS per second | | |
|---|---|---|---|---|---|---|---|---|---|
| | | | OLS | slice | monomvn | Gibbs | slice | monomvn | Gibbs |
| Horseshoe | 20 | 200 | 26.93% | 14.67% | 14.52% | 14.53% | 1626 | 7281 | 8395 |
| | 100 | 1000 | 16.47% | 6.06% | 6.04% | 6.03% | 387 | 747 | 792 |
| | 200 | 2000 | 14.58% | 4.54% | 4.54% | 4.54% | 203 | 183 | 187 |
| | 500 | 5000 | 10.05% | 2.61% | 2.61% | 2.62% | 57 | 16 | 17 |
| | 1000 | 10000 | 6.85% | 1.64% | 1.64% | 1.64% | 36 | 4 | 4 |
| Laplace | 20 | 200 | 27.27% | 16.31% | 16.25% | — | 2357 | 12875 | — |
| | 100 | 1000 | 17.06% | 7.21% | 7.15% | — | 573 | 1257 | — |
| | 200 | 2000 | 14.56% | 5.29% | 5.20% | — | 365 | 306 | — |
| | 500 | 5000 | 10.01% | 3.13% | 3.10% | — | 84 | 27 | — |
| | 1000 | 10000 | 6.77% | 1.95% | 1.94% | — | 38 | 5 | — |
| Ridge | 20 | 200 | 27.36% | 17.33% | 17.34% | — | 2399 | 22608 | — |
| | 100 | 1000 | 16.90% | 8.50% | 8.75% | — | 669 | 1668 | — |
| | 200 | 2000 | 14.38% | 6.42% | 6.69% | — | 342 | 362 | — |
| | 500 | 5000 | 9.90% | 4.18% | 4.40% | — | 89 | 30 | — |
| | 1000 | 10000 | 6.85% | 2.93% | 3.09% | — | 38 | 6 | — |

Table 11: This table compares the error and effective sample size (ESS) per second for various sampling algorithms. The regressors are generated with factor structure, with blocks of consecutive 20 variables having correlation 0.8 but independent from the other blocks. The response variable is generated according to the sparse Gaussian model described in the text. The column "Gibbs" denotes our own implementation of the standard Gibbs sampler for the horseshoe prior. In this table, generated on a machine running the MKL linear algebra library, `monomvn` achieves very high ESS per second when the data size is small, but is still less efficient than the elliptical slice sampler as sample size and dimensionality grows; see the discussion on parallelization in section 2.4.